\documentclass[fleqn,twoside]{article}
\usepackage{espcrc2}

\usepackage{graphicx}
\usepackage{epsfig}

\newcommand{\AmS}{{\protect\the\textfont2
  A\kern-.1667em\lower.5ex\hbox{M}\kern-.125emS}}

\newcommand{\QQ}{\mbox{$Q^2$}}

\newcommand{\eTj}{\mbox{$E_T^{\,\mathrm{jet}}$}}

\title{Virtual photon structure at HERA}

\author{P. J. Bussey\address
       {Department of Physics and Astronomy, University of Glasgow, 
        Glasgow G12 8QQ, United Kingdom\protect\\[3mm]
        for the H1 and ZEUS Collaborations}}
       
\begin{document}

\begin{abstract}
An overview is given of the ongoing measurement at HERA of the parton
structure of the photon, as a function of its virtuality.  Preliminary
ZEUS results show disagreement  with an NLO QCD calculation.
\vspace{1pc}
\end{abstract}
\maketitle

\section{INTRODUCTION}
In $ep$ collisions at HERA, the incoming electron or positron 
radiates a virtual vector boson which then interacts with the proton.
In the hardest high-energy collisions, this 
boson may be a $W$ or $Z$, but the vast majority of collisions 
are mediated by a virtual photon.  Its
virtuality \QQ\ varies from a lower limit of  
$O(10^{-13}$ GeV$^2)$, given by the mass squared of the electron,
to an upper limit of $O(10^{4}$ GeV$^2)$ given by 
the energy of the collider.  The large majority of collisions
are at very low \QQ\ values, where the photon is quasi-real, and are
referred to as photoproduction reactions. The highest-\QQ\ range is that 
of deep inelastic scattering (DIS).  At intermediate \QQ\ values 
of the order of 1 GeV$^2$, there is a transition region.

While all interactions of the photon at HERA are assumed to be
governed by QCD, the photoproduction and DIS regions exhibit different
features.

The quasi-real photon may interact in two basic ways: it may couple
immediately to a hard quark-antiquark pair, or it may interact by
means of an intermediate partonic structure.  The latter process, for
quasi-real photons, is often modelled by means of the vector meson
dominance hypothesis, in which the photon fluctuates into a mesonic
state which then interacts with the proton.  Hard scatters may then
occur in which a parton originating from this state scatters off a
parton from the proton.  This scatter may be calculated in perturbative
QCD, but the original fluctuation of the photon into a partonic state
is not necessarily calculable in this way.  Hence the
probability $f_{a,\gamma}$ that a parton $a$ is found with a fraction
$x_\gamma$ of the photon energy must often be obtained phenomenologically.
Models based on a hadronic structure have been used, and also models
based on fits to experimental data from photon-photon scattering.  

In DIS, the partonic structure of the interacting photon
is often taken to be entirely calculable in perturbative QCD.  The 
non-perturbative component to the photon structure will then  
die away with \QQ\ through the transition region mentioned above. This region  
corresponds to the mass range of the vector mesons and
the range below which perturbative methods in QCD may be expected to fail. 

In this report, recent preliminary measurements by ZEUS of the resolved 
structure of the virtual photon are discussed.  They are made over a 
wide \QQ\ range in an attempt to understand better the 
partonic behaviour of the photon.

\section{TYPES OF PHOTON PROCESS}

\begin{figure}[t!]
\vspace*{-0.5ex}
\hspace*{20mm}\raisebox{0mm}{\epsfig{file=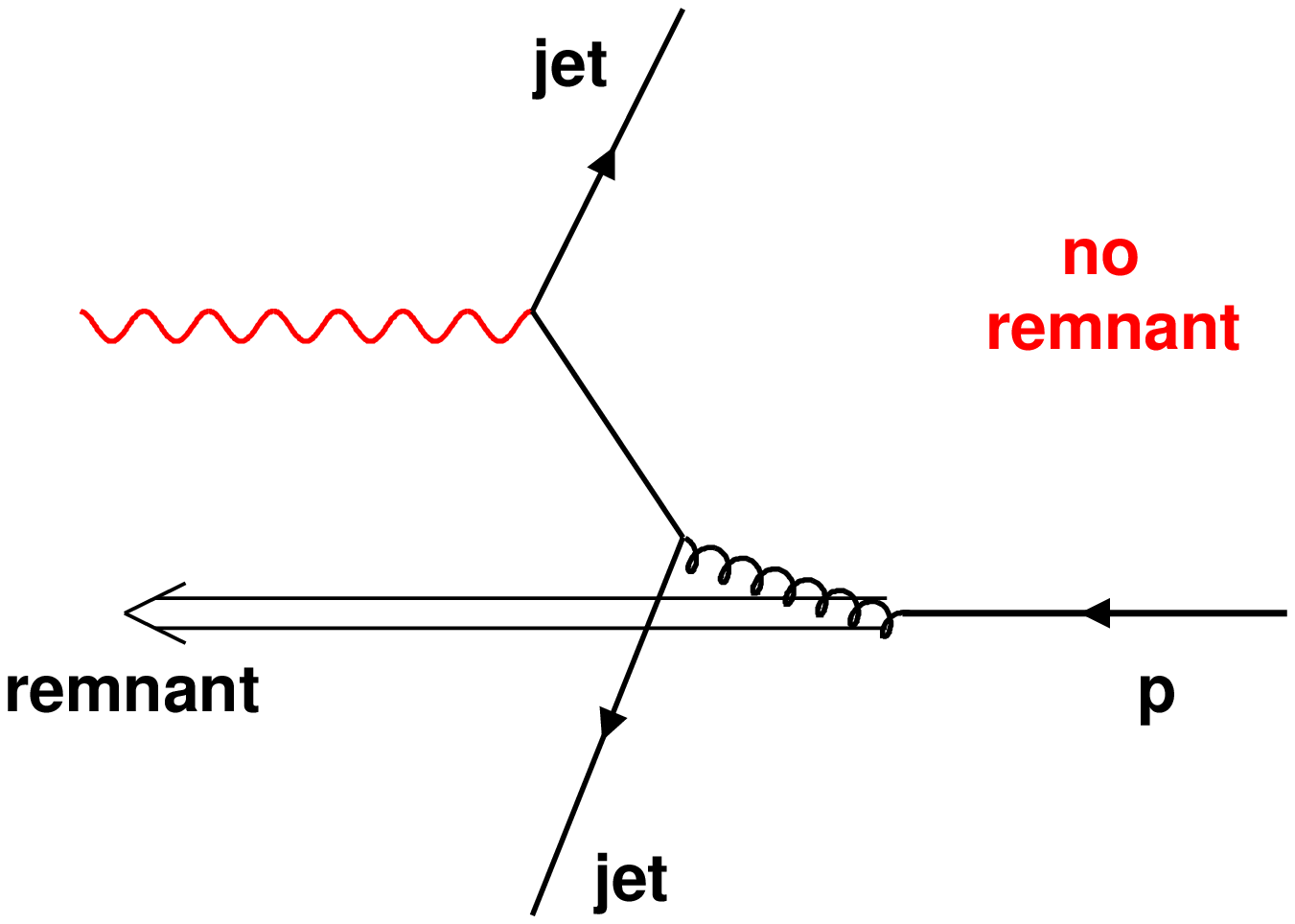,width=4.6cm,%
bbllx=120pt,bblly=90pt,bburx=520pt,bbury=400pt,clip=yes}}
\hspace*{1cm}\raisebox{10mm}{(a) Direct}\\
\hspace*{15mm}\raisebox{0mm}{\epsfig{file=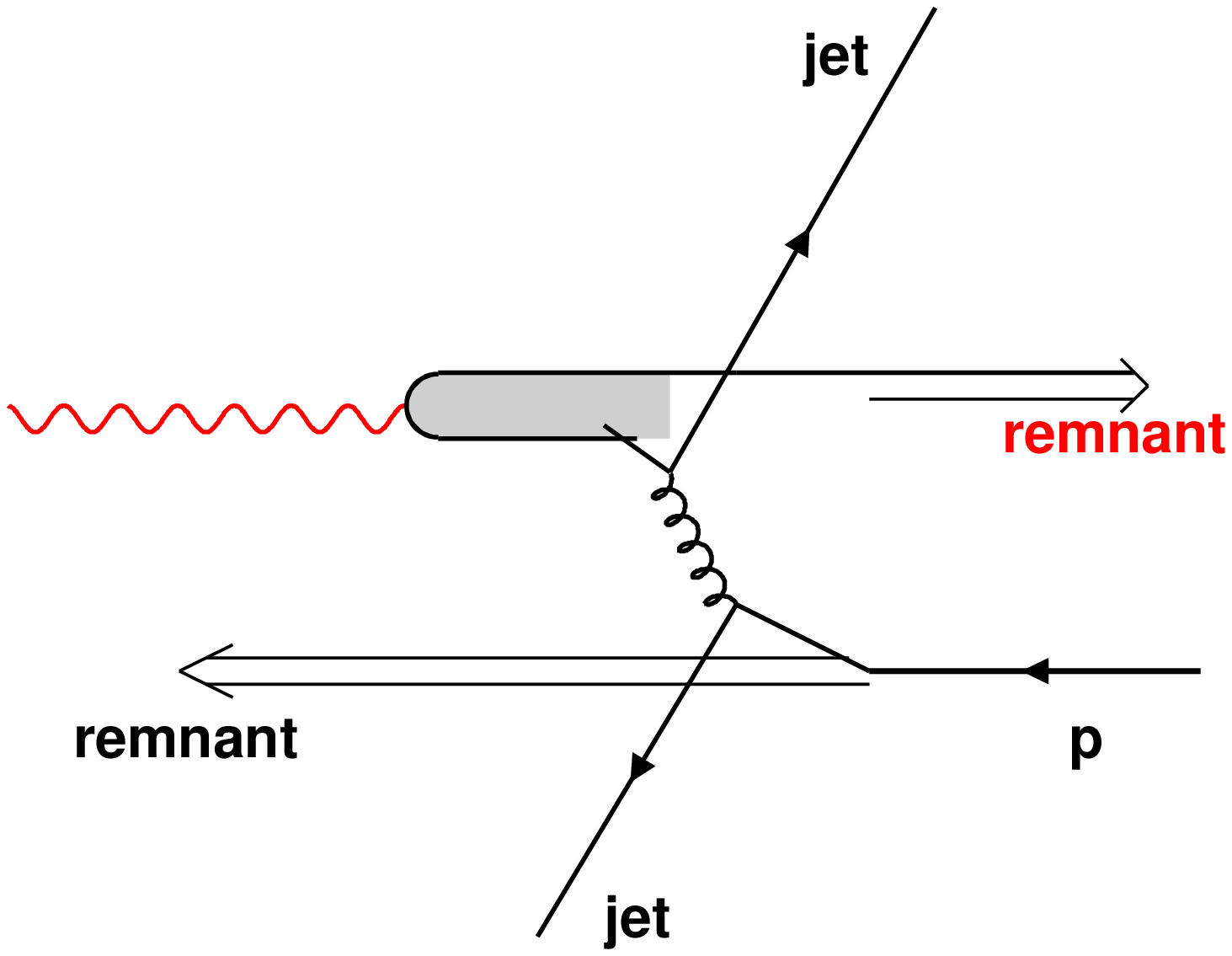,width=5.0cm,%
bbllx=100pt,bblly=90pt,bburx=520pt,bbury=400pt,clip=yes}}
\hspace*{1cm} \raisebox{10mm}{(b) Resolved} \\[-3ex]
\hspace*{20mm}\raisebox{0mm}{\epsfig{file=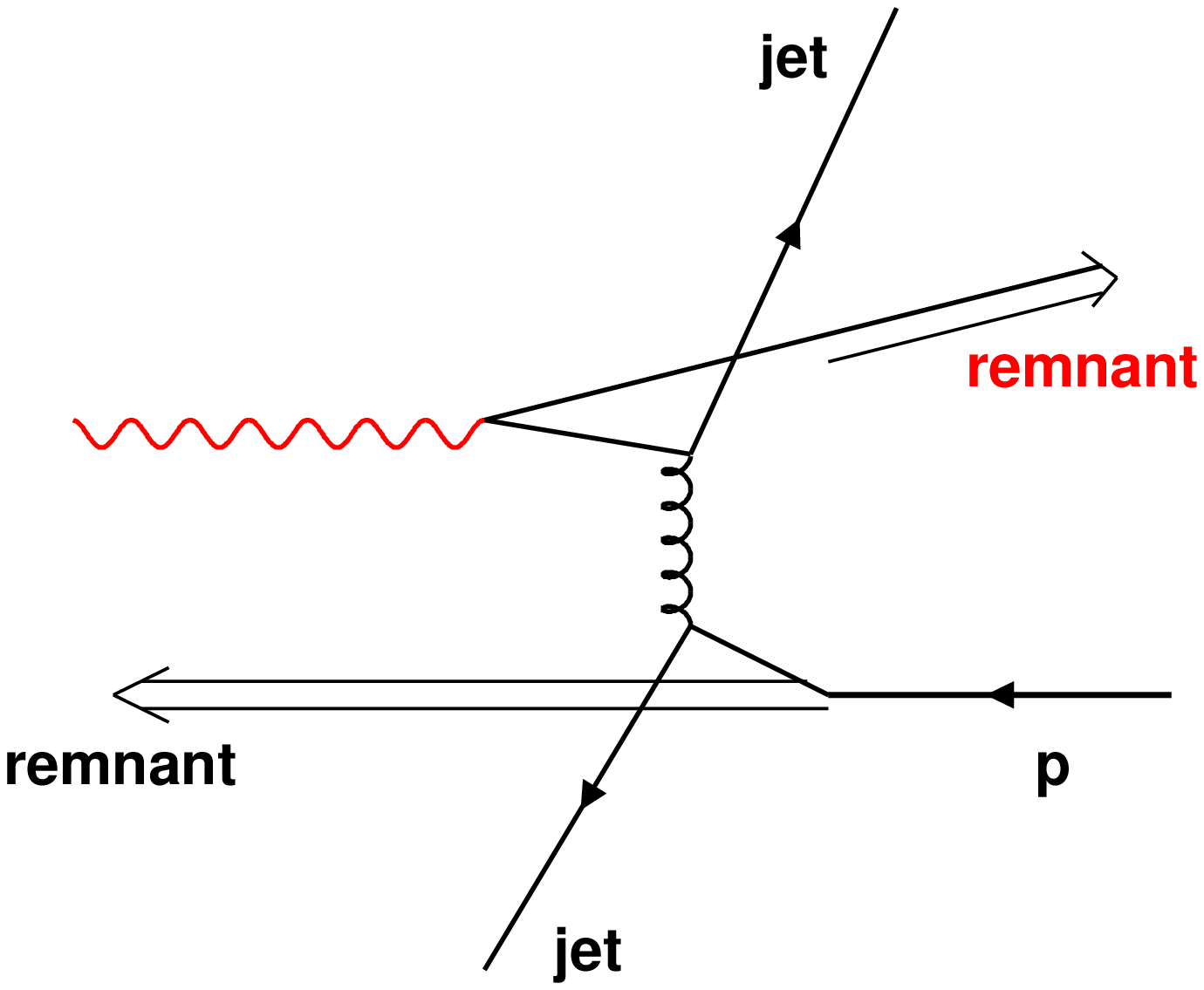,width=4.4cm,%
bbllx=120pt,bblly=90pt,bburx=520pt,bbury=430pt,clip=yes}}
\hspace*{1cm} \raisebox{9mm}{(c) Perturbative resolved}\\[-11ex]
\caption{Examples of photon interactions. }
\end{figure}

The basic types of hard photon process are illustrated in Fig.\ 1.  In
direct processes (a) the entire photon takes part in the hard
interaction and there is no remnant in the photon direction.  In
resolved processes such as (b), the photon fluctuates into a state out
of which a parton emerges to scatter off a parton from the proton.
There is a photon remnant which takes the fraction of the photon
energy which is not given to the hard scatter. This diagram indicates 
processes in which the partonic state is non-perturbative in nature,
and sometimes though controversially termed `hadronic'. The hard
scatter itself, which produces the final-state jets, is of course
calculable in perturbative QCD.

As depicted, (a) and (b) are lowest-order QCD diagrams with regard to the
hard scatter, and the direct and resolved classes are separate.
Corresponding diagrams at higher order can be drawn.  While it is
still often possible, for calculational purposes, to divide these into
`direct' and `resolved' classes, more care is needed and the division 
between the two classes can become somewhat arbitrary.

Diagram (c) indicates a process in which the photon splits into a
$q\bar q$ pair, and which is NLO compared to (a). When this vertex involves
a sufficiently high momentum transfer it is perturbatively calculable,
as is the case if the $q\bar q$ pair have a high $p_T$ or if the
incoming photon is highly virtual.  A photon remnant is present,
though not necessarily at low $p_T$, and only part of the photon
energy is given to the final-state jets. This type of process may therefore be
classed as resolved. In deciding whether to call the remaining photon
system a remnant or a third jet, however, an arbitrary decision must be taken
as to the $p_T$ value that will constitute the boundary between the
two cases.

The basic formula for calculating dijet cross sections
in $ep$ collisions can be written as
\begin{eqnarray*}
\lefteqn{d\sigma_{ep\to e+\mathrm{jets}} = }  \\
  & & \sum_{a,b}\int_0^1{dy\,f_{\gamma^*/e}(y,Q^2)} \\ & & \hspace*{3ex}
\times \int_0^1{dx_{\gamma^*}\,f_{a/\gamma^*}(x_{\gamma^*},Q^2,\mu^2_{F\gamma^*})}
\\  & & \hspace*{5ex}
\times\int_0^1{dx_p\,f_{b/p}(x_p,\mu^2_{Fp})\,d{\sigma}_{ab\to \rm{jets}}(\mu_R).
}
\end{eqnarray*}
Here $f_{\gamma^*/e}(y,Q^2)$ denotes the probability for the incoming $e$ 
to radiate a given virtual photon, and $f_{a/\gamma^*}$ and $f_{b/p}$
denote the parton density functions (PDFs) of the photon and proton.
The hypothesis of factorisation asserts that these are process-independent.
For direct processes, $f_{a/\gamma^*} = \delta(x_{\gamma^*} - 1)$.
In general there are two contributions to the resolved ${\gamma*}$ PDFs:
$$
f_{a/\gamma^*}
  = f_{a/\gamma^*}^{\mathrm{nonpert}} +f_{a/\gamma^*}^{\mathrm{pert}} 
$$
The boundary between the non-perturbative and perturbative regions
corresponds to the so-called `factorisation scale' $\mu_F^2$. The
renormalisation scale $\mu_R^2$ refers to the momentum transfer
squared of the hard QCD scatter.  In DIS,
$f_{a/\gamma^*}^{\mathrm{nonpert}}$ is often assumed to be zero.

\newpage

\section{DIJET EVENTS}

In their recent analysis, ZEUS identify dijet events in the rear
($\gamma^*$) hemisphere of the $\gamma^* p$ centre-of-mass frame with
\eTj$> 7.5,\, 6.5$ GeV.  A measure of $x_\gamma$ is 
$$ x_\gamma^{\;\mathrm{obs}} =
\left(E_T^{\;1} e^{-\eta_1} +
E_T^{\;2} e^{-\eta_2}\right)/2E_{\gamma^*}  $$
where the jets are labelled 1, 2 and  $\eta$ is pseudorapidity. 
Figure 2 shows a typical $x_\gamma^{\;\mathrm{obs}}$ distribution in photoproduction.
The prominent peak is due largely to direct processes, while
the tail is due largely to resolved processes; its precise shape 
depends on the cuts applied to the observed jets. 

The incoming virtual photon energy $E_{\gamma^*}$ is measured by tagging the 
scattered electron/positron in the ZEUS calorimeter system.
A specially built calorimeter was inserted close to the beam line in order to enable 
\QQ\ values through the transition region to be studied. 

\begin{figure}[t]
\vspace*{0.5ex}
\epsfig{file=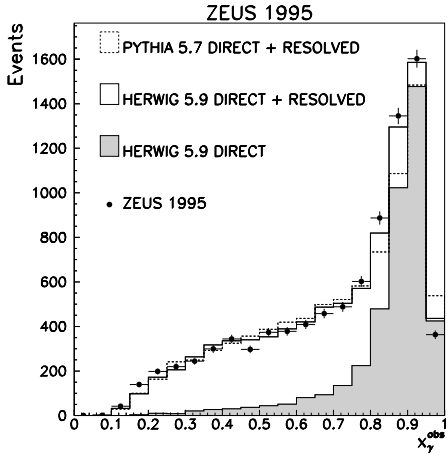,width=6cm}\\[-7ex]
\caption{Typical distribution $x_\gamma^{\;\mathrm{obs}}$ 
observed in photoproduction at HERA.  Calculated contributions are indicated  
from the resolved and direct processes at LO in the QCD scatter,
with added initial and final state radiation.}
\end{figure}

A pure LO direct process has $x_\gamma$ = 1, but the effects of
initial and final state radiation and hadronisation reduce the value
of the observed quantity.  It is nevertheless useful to regard the
regions $x_\gamma^{\;\mathrm{obs}} < 0.75 $ and
$x_\gamma^{\;\mathrm{obs}} > 0.75$ as resolved-dominated and
direct-dominated respectively.  Provided the proton PDFs are known,
the direct processes are reliably calculable in perturbative QCD, both
in photoproduction and DIS. The present investigation is concerned
primarily with the size of the resolved-dominated contribution as a
function of \QQ. This can be studied by means of the ratio $$ R=
\frac{d\sigma}{dQ^2}(x_\gamma^{\;\mathrm{obs}} < 0.75)\left/
\frac{d\sigma}{dQ^2}(x_\gamma^{\;\mathrm{obs}} > 0.75)\right. 
$$
%$$ R= \frac{\Hl\frac{d\sigma}{dQ^2}(x_\gamma^{\;\mathrm{obs}} < 0.75)}{
%\Hi\frac{d\sigma}{dQ^2}(x_\gamma^{\;\mathrm{obs}} > 0.75)}
%$$
A number of  experimental and theoretical uncertainties tend to cancel in
evaluating $R$, whose variation with the kinematics of the
process may be used to study the behaviour of the effective partonic
structure of the virtual photon.

The DISASTER++ Monte Carlo (D. Graudenz) was used to calculate parton
cross sections at NLO.  These were corrected to hadron level using the
well-tried ARIADNE program. The data were corrected to hadron level
using PYTHIA.

\section{RESULTS} 
The measured cross section for the production of dijets, within the
given kinematic acceptance, as a function of \QQ, is shown in Fig.\ 3.
The data are compared to DISASTER calculations using a renormalisation
scale $\mu_R^2= Q^2+E_T^2$, with $\mu^2_{Fp}$ also set to this
value. The dominant theoretical uncertainty was estimated by halving
and doubling this value.  DISASTER cannot be reliably used below a
\QQ\ value of approximately 2 GeV$^2$.  It is of interest to see that
DISASTER predicts the direct-dominated cross sections excellently, but
under\-esti\-mates the total cross section through a serious
underprediction of the resolved-dominated component.  A standard PDF
is used for the proton, the results being insensitive to this choice.
DISASTER evaluates only point-like photon processes,  namely
the direct and the perturbative resolved components.  No
non-perturbative photon PDF or photon factorisation scale therefore
applies in this calculation. 

A similar situation is observed when the cross sections are plotted as a
function of the transverse jet energy \eTj. The underestimation of 
the cross sections is substantial and shows no strong \eTj-dependence.  As
a further study, cross sections have been plotted as a function of the
pseudorapidity of the more forward jet in the proton direction.
DISASTER was also evaluated using $\mu_R^2= Q^2$, although this is not a
plausible choice for the present \eTj\ values unless $\QQ \gg
(\eTj)^2$. It does, however, illustrate the sensitivity
of the calculations to this scale.

\begin{figure}[t!]
\vspace*{-2ex}
\epsfig{file=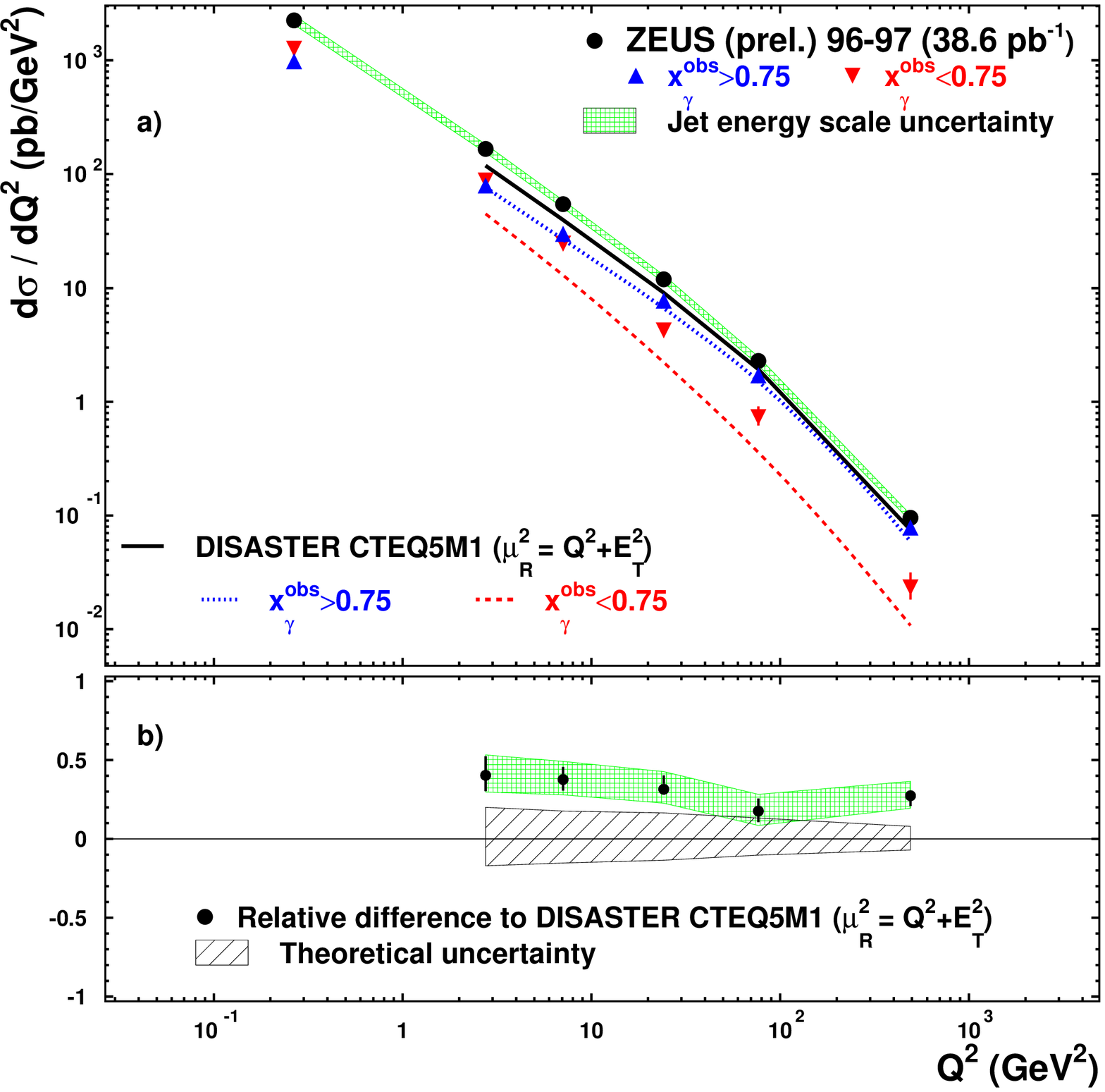,width=7.4cm}\\[0ex]
\epsfig{file=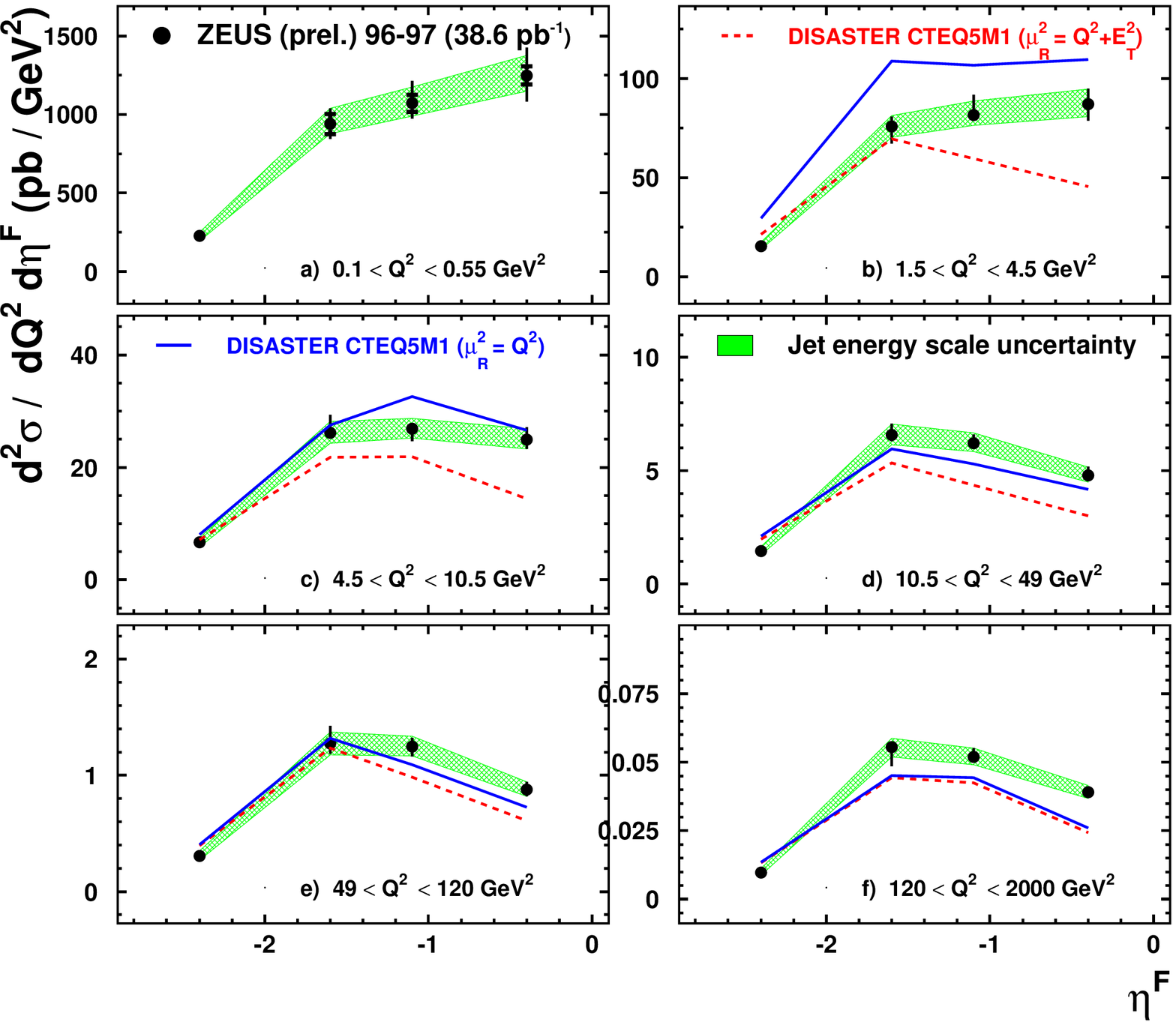,width=7.4cm}\\[-7ex]
\caption{Upper plots: dijet cross sections as a function of \QQ\ of
the virtual photon.  Comparison is made with predictions from
DISASTER for the direct and resolved dominated regions.  Lower
plots: cross sections as a function of pseudorapidty of the
forward jet, for different \QQ\ regions.  Comparison is made with
predictions from DISASTER using two different renormalisation
scales.}
\end{figure}

Figure 4 shows the ratio $R$, plotted as a function of \QQ\ in three
different (\eTj)$^2$ ranges (a-c).  The value of $R$ in 
photoproduction events (\QQ $\approx 0)$, not plotted,  is estimated to be similar,
within errors, to the  value for the lowest \QQ\ points shown.  $R$
is seen to fall sharply over the transition region for the lowest
\eTj\ range, and more gently for the higher ranges. 
In (d), $R$ is plotted in a different analysis in which dijet events
containing a $D^*$ meson were selected, and no fall in $R$ is evident.
The presence of either high \eTj\ or a heavy quark forces a high
momentum scale on the QCD process.  In such situations, the process is
expected to be perturbatively calculable, with no prima facie reason
for a strong \QQ\ dependence.  The stronger fall-off at low
\eTj\ values is consistent with the presence of strong non-perturbative
photon structure effects in processes involving light quarks.

Nevertheless, even at high \QQ, $R$ does not fall to zero but seems to
level off at a constant value that is not strongly dependent on \eTj.
This may be interpreted in terms of an effective partonic structure to the
photon, even at high virtualities.  The DISASTER calculation, evaluated
using $\mu_R^2= Q^2+E_T^2$, is unable to account for the data, and the discrepancy
of the order of 50\% does not vary strongly with  the event
kinematics. It is evident that the full cross section is not being
correctly simulated by the perturbative model used.  However the
JETVIP model, although including a non-perturbative photon PDF, fares
worse and cannot be considered competitive at present.

In an earlier study, H1 measured dijet events over a range of moderate
\QQ\ values and interpreted the resolved cross sections in terms of an
effective photon parton density (Fig.\ 5). This at first falls sharply
from its value at $\QQ \approx 0$, but afterwards remains constant
within statistics.  However, it was not possible to demonstrate a
significant deviation from the falling NLO QCD calculation. The
present ZEUS results appear to lend weight to the view that the
effective parton structure of the photon remains approximately
constant over this \QQ\ range.

\begin{figure}[b!]
\vspace*{-9ex}
\hspace*{-0ex}\epsfig{file=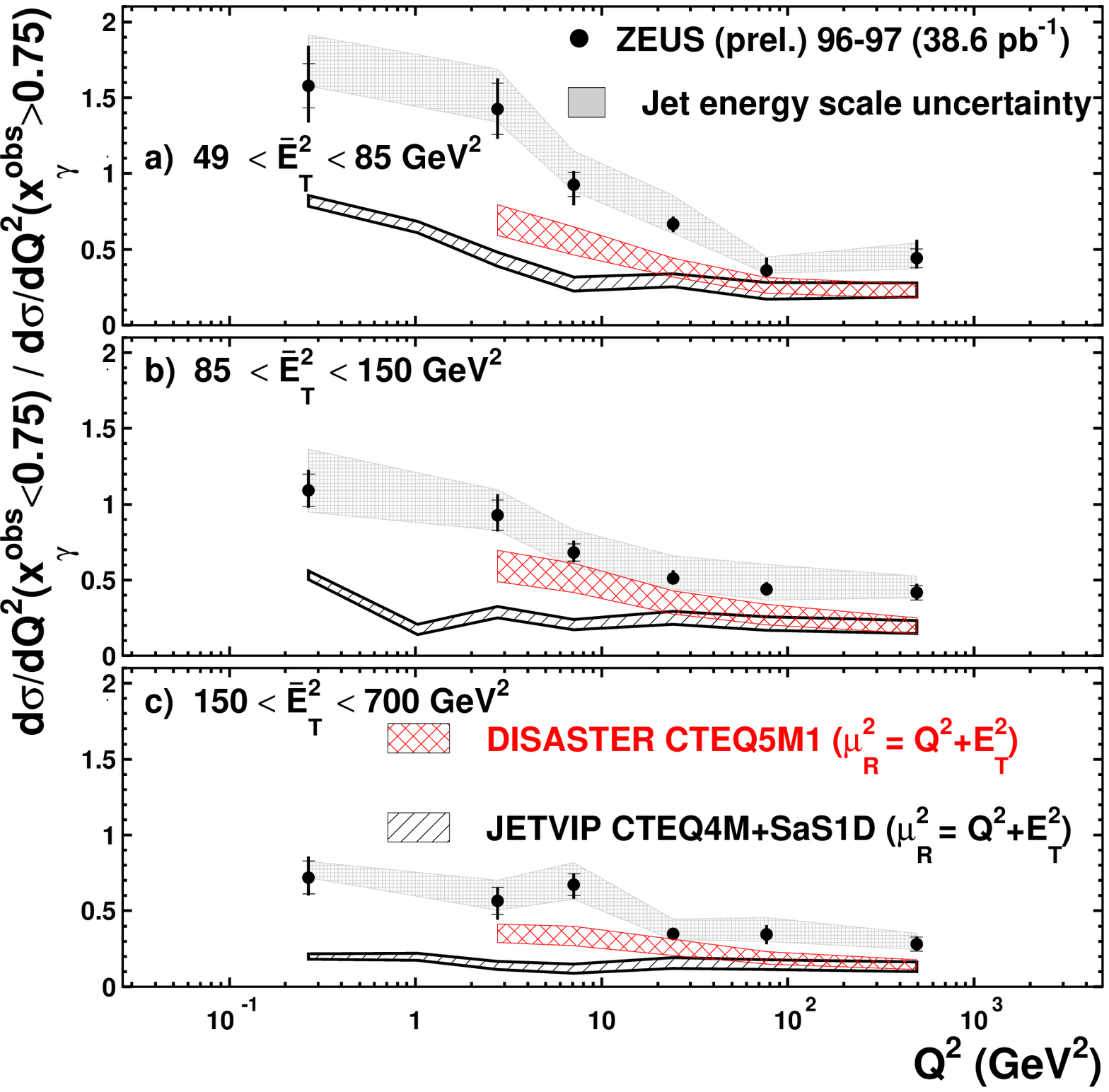,width=8.2cm}\\[2ex]
\hspace*{-0.8ex}\epsfig{file=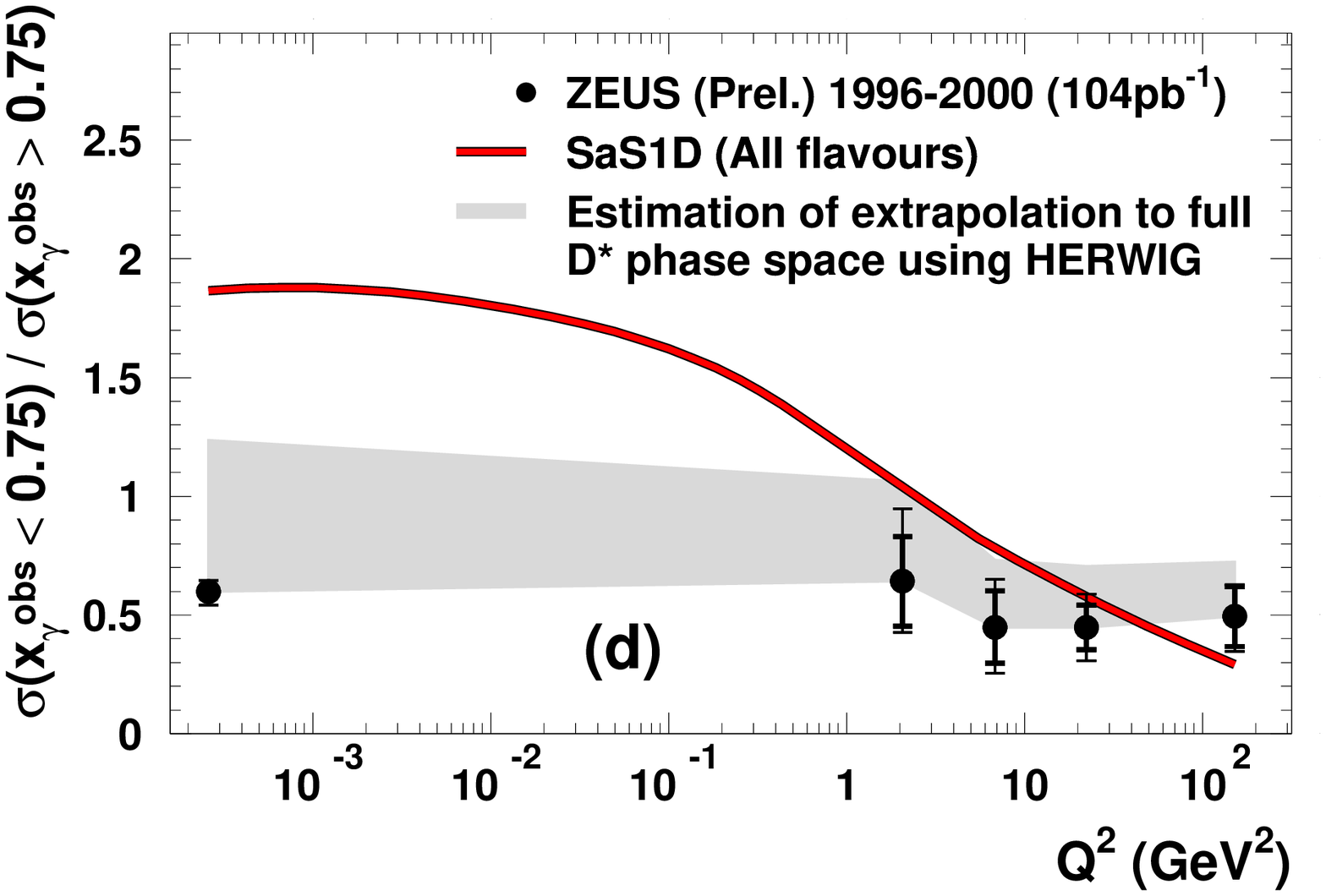,width=8.4cm}
\\[-7ex]
\caption{Cross section ratio $R$  as a function of \QQ\ of
the virtual photon, (a-c) in different ranges of (\eTj)$^2$. Comparison
is made with predictions from DISASTER and JETVIP. (d) $R$
for events containing an identified D$^2$
meson.}
\end{figure}

\section{CONCLUSIONS}

ZEUS has presented preliminary measurements of dijet cross sections
over a range of \QQ\ values crossing the transition between
photoproduction and DIS.  A ratio of the resolved-dominated to
direct-dominated cross sections is defined. It falls across this
region, as the likely vector-meson dominated region of photon
structure is left behind, but appears to attain a stable non-zero
value.  A component to the cross sections is indicated which is not
modelled in the NLO QCD calculation DISASTER++.  The shortfall is
associated with resolved processes.  Possible explanations might be a
non-perturbative partonic structure in the photon even at high \QQ, or
a surprisingly strong higher order perturbative contribution. The
results are consistent with earlier H1 data.  In events containing a
charmed meson, the ratio does not fall with \QQ, but remains
approximately flat over the entire range, at a similar value to the
overall dijet sample at high \QQ.

\begin{figure}[b!]
\vspace*{-1ex}
\hspace*{-1.5ex}\epsfig{file=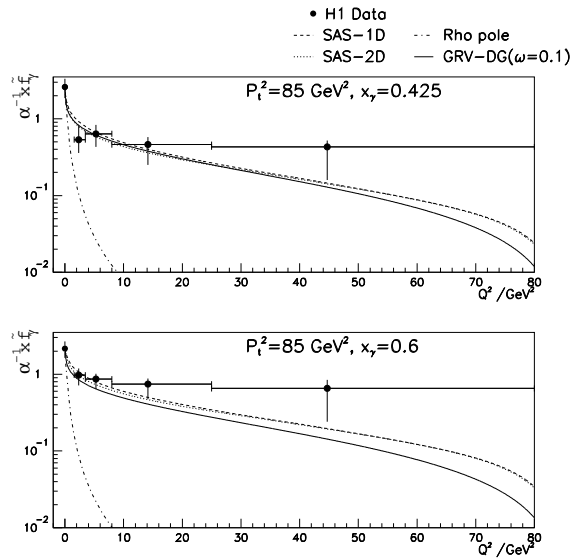,width=8cm}
\\[-7ex]
\caption{Effective parton density in the virtual photon, 
evaluated by H1, as a function of \QQ, compared with NLO predictions
(Eur. Phys. J. C13 (2000) 397) }
\end{figure}

\end{document}